\documentclass{article}
\usepackage{arxiv_style}

\newcommand{\fooAlter}{\hspace{0pt}\textcolor{blue}{$\bullet$} \hspace{5pt}}

\AtBeginDocument{%
  \providecommand\BibTeX{{%
    \normalfont B\kern-0.5em{\scshape i\kern-0.25em b}\kern-0.8em\TeX}}}

\usepackage[utf8]{inputenc} 
\usepackage[T1]{fontenc}    
\usepackage{hyperref} 
\usepackage{scalerel,stackengine}
\newcommand\Tstrut{\rule{0pt}{2.1ex}}       
\usepackage{colortbl}
\usepackage[export]{adjustbox}
\usepackage{graphicx}
\usepackage{xcolor}
\usepackage{subcaption}
\usepackage{colortbl}
\usepackage{bbm}
\usepackage{url}            
\usepackage{booktabs}       
\usepackage{amsfonts}       
\usepackage{nicefrac}       
\usepackage{microtype}      
\usepackage{natbib}
\title{Infrastructure Ombudsman: Mining Future Failure Concerns from Structural Disaster Response}
\author{
Md Towhidul Absar Chowdhury\\
  \small{Rochester Institute of Technology} \\
  \texttt{mac9908@rit.edu} \\
  \And
  Soumyajit Datta$^*$\\
  \small{Maulana Abul Kalam Azad University of Technology}\\
  \texttt{soumyajitdatta123@gmail.com} \\
\And
Naveen Sharma \\
  \small{Rochester Institute of Technology}\\
  \texttt{naveen.sharma@rit.edu} \\
 \And
Ashiqur R. KhudaBukhsh \\
  \small{Rochester Institute of Technology}\\
  \texttt{axkvse@rit.edu} \\
}

\begin{document}

\maketitle
\begin{abstract}


Current research concentrates on studying discussions on social media related to structural failures to improve disaster response strategies. However, detecting social web posts discussing concerns about anticipatory failures is under-explored. 
If such concerns are channeled to the appropriate authorities, it can aid in the prevention and mitigation of potential infrastructural failures. 
In this paper, we develop an \textit{infrastructure ombudsman} -- that automatically detects specific infrastructure concerns. 
Our work considers several recent structural failures in the US. 
We present a first-of-its-kind dataset of 2,662 social web instances for this novel task mined from Reddit and YouTube. 


\end{abstract}

\section{Introduction}

On January 28, 2022, at 6.39 a.m. EST, the Fern Hollow Bridge in Pittsburgh, Pennsylvania collapsed. Due to the timing of the failure, thankfully, fewer vehicles were on the bridge and only ten people were injured with no fatalities. Pittsburgh, also known as the City of Bridges, was getting ready for a visit from President Biden that day. 
Biden visited the collapse site and assured federal assistance to rebuild the bridge on the spot. This infrastructural failure, coinciding with a high-profile political visit and a push towards passing the Build Back Better infrastructure bill, attracted considerable media attention to the flailing infrastructural health in the US. 

As we were sifting through the social web discussions surrounding this issue, broad themes such as words of compassion for the victims and typical responses in social web political discourse such as political name-calling, conspiracy theories, and partisan mud-slinging emerged. However, apart from these expected social web reactions, we noticed a small minority of interactions that talked about anticipatory failures of other bridges in the US. Table~\ref{tab:BridgeReactions} lists a few illustrative examples.      

\begin{table}[t]

{
\small
\begin{center}
     \begin{tabular}{| p{7cm} |}
     \hline
    \Tstrut 
     Social Media Post\\
     
    \hline
    \Tstrut 
    \cellcolor{gray!15}\textit{There is a bridge in Lowell Massachusetts, it goes over the Merrimack river and it is rusted strait through.  It won’t be long before we suffer major injuries because that bridge is always bumper to bumper traffic!}
    \\
     \hline 
     \Tstrut
     \textit{I'm surprised the New Kensington bridge wasn't the first to go. Haha. Terrible condition.}\\
     \hline 
     \Tstrut
     \cellcolor{gray!15}\textit{The bridge on 1-81 that spans the Potomac between WV and MD that could be next. It has a lot of 18 wheelers beating it up.  I kept a hammer in my car to get out if it collapsed when I was on it.}\\
     \hline 
    \end{tabular}
\end{center}
\caption{Illustrative examples indicating concerns over other vulnerable bridges following the collapse of the Fern Hollow Bridge in Pittsburgh.}
\label{tab:BridgeReactions}}
\end{table}

The comments in Table~\ref{tab:BridgeReactions}, if mined efficiently and surfaced to the appropriate authorities, can present an effective path to intercept structural failure concerns. Failures that are yet to happen - but a responsible citizen is worried that they might. Vulnerabilities that they perhaps noticed before, but the sudden, exogenous shock in the form of a structural failure gives them an outlet to voice their concerns. Extant research on disaster response discourse focuses on a diverse set of tasks that include: efficient distribution of relief~\cite{varga-etal-2013-aid}, crisis management, handling emergencies etc \cite{horita2017bridging}. Much of this research has focused on natural disasters such as typhoons \cite{zou2023social}, earthquakes \cite{sakaki2012tweet}, floods \cite{feng2020flood}, and accidents with severe fatalities \cite{liu2020impact}. Analyzing social media responses to structural failures such as bridges or building collapses is rather new and looking beyond the immediate, and focusing on future potential crises has no prior literature to the best of our knowledge.

This paper presents \textit{infrastructure ombudsman}, an automated social media listener tool that surfaces infrastructure concerns. Via a curated corpus of 2,662 instances (271 positives and 2391 negatives)\footnote{https://github.com/towhidabsar/InfrastructureOmbudsman}, we demonstrate that state-of-the-art NLP and AI methods can be harnessed to build such tools that can aid humans in identifying infrastructure concerns. While the Pittsburgh Bridge collapse was an exogenous shock that got people discussing other potential vulnerabilities, it is possible that our findings point to a broader human pattern where structural collapses trigger similar thoughts about anticipatory failures. Our dataset thus considers several recent infrastructure failures in the US and combines effective rare-class mining methods to present a more holistic resource to this novel task.

Our contributions are the following.

\noindent\fooAlter\textbf{\textit{Novel task: detecting infrastructure concerns:}} We define a new task of detecting infrastructure concerns from social web posts. Our task presents a marked departure from existing disaster response literature where our goal is to identify citizen concerns about possible, future structural failures.   

\noindent\fooAlter\textbf{\textit{New resource:}} We release a dataset of  2,662 instances (271 positives and 2,391 negatives) of infrastructure concerns mined from millions of Reddit posts and YouTube comments. We present a suite of strong baselines trained on this dataset and demonstrate the feasibility of automatically detecting infrastructure concerns with reasonably high precision and recall. On unseen data, we demonstrate that our content classifier can effectively aid humans in detecting potential candidates that express infrastructure concerns. 

\begin{table*}[htb]

{
\small
\begin{center}
     \begin{tabular}{| p{11cm}  | p{2cm} |}
     \hline
    \Tstrut 
     Social Media Post & Label\\
     
    \hline
    \Tstrut 
    \cellcolor{blue!15}There is a bridge in Lowell Massachusetts, it goes over the Merrimack river and it is rusted strait through.  It won’t be long before we suffer major injuries because that bridge is always bumper to bumper traffic!&\cellcolor{gray!15} Positive
    \\
     \hline 
  \Tstrut 
    \cellcolor{blue!15}Pennsylvania has some of the worst roads in the country. As soon as you cross the state line from New York state to PA state the roads immediately are filled with potholes and are built with shitty rugged concrete&\cellcolor{gray!15} Positive
    \\
     \hline 
     \Tstrut
     \cellcolor{red!15}There’s a railroad overpass not far from where I live. About four years ago myself and my daughter were stopped waiting for a car to go through, it’s a one lane back road and only one vehicle at a time can drive through. We both noticed that there was a large crack, at least a foot long and maybe two or three inches wide? \ldots & \cellcolor{gray!15} Negative\\
     \hline 
     \Tstrut
     \cellcolor{red!15}Man, here in southeast ct theyre sandasting and repainting every bridge every couple years. Its crazy watching how fast they rust back up over the years. Ive seen them hit the same one twice enough years later. I would assume our bridges would collapse too if they didnt do this regularly. Some states are just stright up neglectful & \cellcolor{gray!15} Negative\\
     \hline

    \end{tabular}
\end{center}
\caption{Illustrative examples highlighting subtleties of our task.}
\label{tab:separability}}
\end{table*}

\section{The Needle and the Haystack}

\subsection{Needle}

Intuitively, users expressing similar concerns following a structural failure is a rare event. But what kind of concerns would be useful to surface to appropriate authorities? Broad, generic concerns (e.g., \textit{Eventually, all US bridges will go down given they are so old}), while helpful to inform overall government priorities, specific, actionable concerns are what we are looking for in this task. Precise location information will be key for the concern to be actionable. 

Table~\ref{tab:separability} presents a few illustrative examples to elucidate this nuance. Unless the content is geo-tagged, it is impossible to infer the location from \textit{There's a railroad overpass \textbf{not far from where I live}}. In contrast, the post about the bridge over the Merrimack River in Lowell Massachusetts uniquely identifies the bridge~\footnote{As per bridgemeister.com, only four bridges go over the Merrimack River of which only one is in Lowell.}. Similarly, the complaint about roads in Pennsylvania has a reasonably precise location close to the state line coming from NY. Our final example in Table~\ref{tab:separability} has precise location information and talks about infrastructural issues. However, it does not qualify for what we are looking for as the post is praising the administration for doing a responsible job rather than worrying about a possible failure.

\subsection{Haystack}

To provide a better sense of the \textit{haystack}, we present a characterization of the social web discourse following the Pittsburgh bridge collapse. We randomly selected 100 YouTube comments with the keyword \texttt{bridge} on videos relevant to the Pittsburgh bridge collapse. Filtering by keywords is an established high-recall 
approach to obtain relevant discussion~\cite{HaltermanKSO21, IJCAI2022Police}. We next characterize these comments through an open-coding approach. We broadly observe the following categories: partisan blame game; conspiracy theory; relief over no casualty; critical of US infrastructure; critical of local administration; sarcasm; and infrastructure concern. Table~\ref{tab:haystack} summarizes the distribution with illustrative examples.  Note that, we do not intend
these categories to be formal or exhaustive, but rather to be illustrative of the broad themes present in the discourse following this incident and provide a rough estimate of the relative distribution of
these themes to underscore that infrastructure concerns are indeed a rare class.

\begin{table*}[htb]
     \begin{tabular}{|p{3cm}|p{1cm} |p{9cm}|}
     \hline
    Category & Count & Social Media Post   \\
     
    \hline 
    Partisan blame game & 29 &\cellcolor{gray!15}\textit{This is why the dum republicans need to approve the Biden relief package and help build the bridges and airports. Republicans need to wake up and smell the coffee.}
    \\
    \hline 
    Conspiracy theory & 17 &\textit{Someone really should check the area for explosives residue this does not seem like an accident it just so happened the president was coming to the exact town a big bridge collapsed in smells kinda fishy}
    \\
    \hline
    Relief over no casualty & 12 &\cellcolor{gray!15}\textit{Could u imagine driving on the bridge when it collapsed    Thank goodness no one died.}
    \\
    \hline
    Critical of US infrastructure & 15 &\textit{The US is also literally falling apart. A couple of trillion won't fix roads, bridges, airports, water infrastructure etc. and stop the decay}
    \\
     \hline
    Critical of local administration & 8 &\cellcolor{gray!15}\textit{Why has this bridge been left in poor condition for so long? Who in Pittsburgh and/or Pennsylvania is responsible for maintaining these structures? How much money have they spent, if any, to maintain and repair these bridges? Why are local and state elected officials and agencies not doing more?
\ldots } \\
    \hline
    Sarcasm & 13 &\textit{Don't worry everyone, nobody will be fired over this. It's perfectly normal for bridges to just fall the Fck down.}
    \\
    \hline
    Infrastructure concern & 2 & \cellcolor{gray!15}\textit{Bay bridge in Maryland is next. Christ, that thing is rickety}
    \\
    \hline
    Others & 4 & 
    \\
    \hline
    \end{tabular}
\caption{Illustrative examples highlighting the diversity in discourse following the Pittsburgh bridge collapse.}
\label{tab:haystack}
\end{table*}

\section{Dataset}
\begin{figure*}[t]
    \centering
    \includegraphics[width=0.9\linewidth]{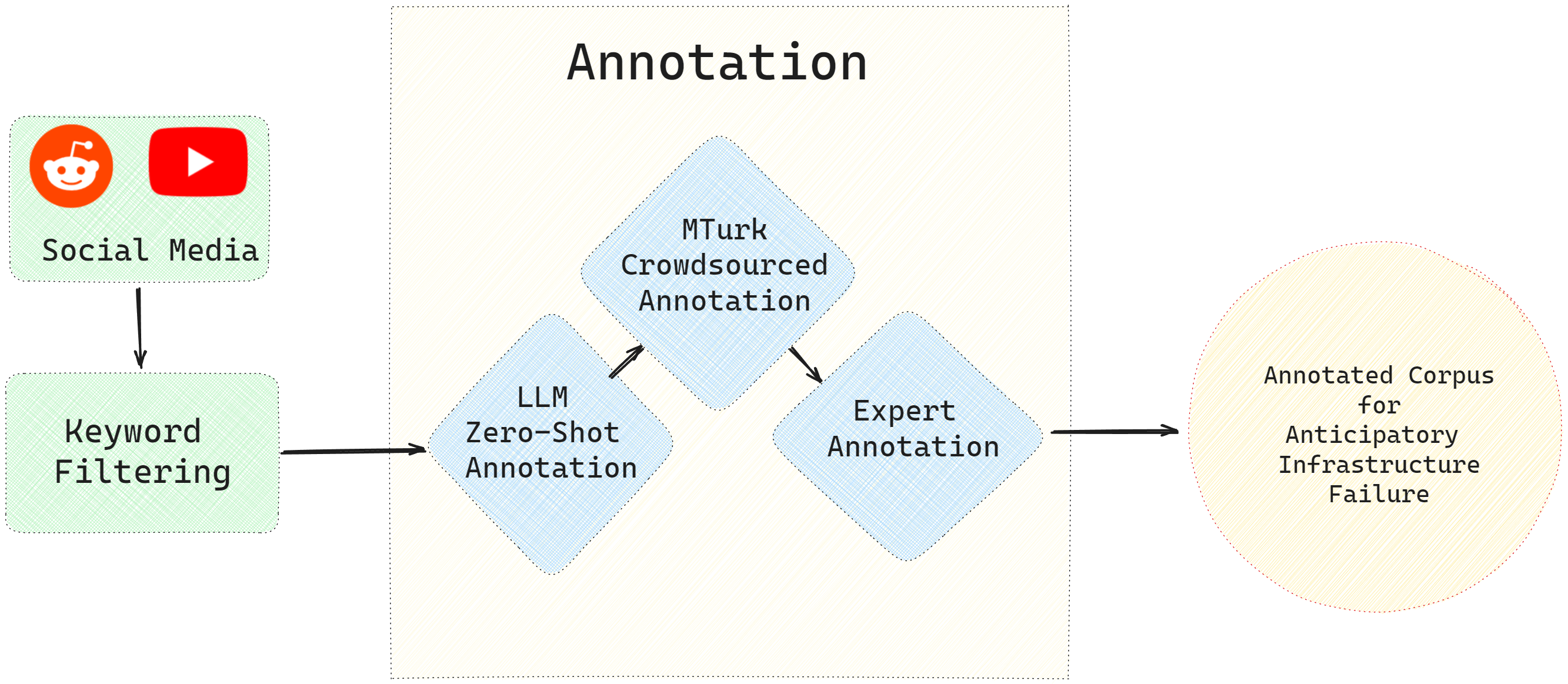}
    \caption{Dataset Creation Pipeline}
    \label{fig:dataset_creation_pipeline}
\end{figure*}
    %
\subsection{Platforms: Reddit and YouTube}
To compile a diverse corpus with broad participation, we select a multi-platform approach. We consider two well-known social web platforms: Reddit and YouTube.
Reddit provides a wide range of subreddit forums catering to different political affiliations and demographics, allowing us to gather infrastructure-related discussions from communities with varied ideological leanings. For YouTube, we consider a longitudinal dataset of 110 million comments on the official handles of three major US cable news networks: CNN, Fox News, and MSNBC (denoted by $\mathcal{D}_\textit{politics}^{YT}$).
This dataset is considered a reliable and diverse snapshot of US political discourse~\cite{khudabukhsh2021we,yoo2023auditing,emnlp2023}. In addition, we also search YouTube with a set of keywords (described next) using publicly available API and collect comments on these videos ($\mathcal{D}_\textit{targeted}^{YT}$). By consolidating data from two different platforms, we seek to build a corpus covering perspectives across the political spectrum, facilitating a more balanced analysis that is less skewed by any individual platform's biases.

Figure~1 presents a high-level, schematic diagram of our data curation pipeline. In what follows, we present a brief description of each step. 

\subsection{Keyword Filtering}


We curate the corpus to include discussions and media content related to several specific search keywords:\\
\{\textit{train derailment}, \textit{infrastructure},
\textit{infrastructure collapse},
\textit{infrastructure concern},
\textit{Ohio train derailment},
\textit{Missouri train derailment},
\textit{Champlain Towers South collapse},
\textit{AdventHealth Orlando parking garage crane collapse},
\textit{Charlotte scaffolding collapse},
\textit{Pittsburgh bridge collapse},
\textit{Fern Hollow Bridge Collapse},
\textit{I-85 Overpass collapse}\}

Keyword-based filtering is a well-established, high-recall approach to pruning datasets tailored for specific interests~\cite{DemszkyGVZSGJ19,HaltermanKSO21,IJCAI2022Police}. For $\mathcal{D}^\textit{Reddit}$ and $\mathcal{D}_{targeted}^\textit{YT}$, the keyword set serves as search keywords. For $\mathcal{D}_{politics}^\textit{YT}$, we consider comments on videos where at least one of the keywords is present in the video title or description.   

Our keyword set captures a broad spectrum of incidents ranging from structural failures such as building collapses to railroad failures such as train derailments. By including specific incidents such as the Champlain Towers South collapse and the AdventHealth Orlando parking garage crane collapse, the corpus allows for in-depth case studies and comparative analyses, shedding light on the unique challenges and responses associated with different types of infrastructure failures.

To summarize, our initial corpus consists of the following sub-corpora: $\mathcal{D}^{Reddit}$ (140,326 posts); $\mathcal{D}_{\textit{politics}}^{YT}$(405,758 YouTube comments); and $\mathcal{D}_{\textit{targeted}}^{YT}$(10,251 YouTube comments). 
We set aside 10,000 posts from $\mathcal{D}^{Reddit}$ and $\mathcal{D}_{\textit{targeted}}^{YT}$ for in-the-wild evaluation (denoted by $\mathcal{D}_\textit{in-the-wild}$).  


\subsection{Harnessing LLMs and Machine Annotation}
\subsubsection{Textual Entailment}
Textual entailment, also known as natural language inference (NLI)~\cite{maccartney2008modeling}, has found recent applications in a broad range of social inference tasks that include (1) estimating media stance on police portrayal~\cite{HaltermanKSO21,IJCAI2022Police}; (2) detecting COVID-19 misinformation~\citep{hossain-etal-2020-covidlies}; and (3) aggregating social media opinions on election fairness~\citep{Capitol2022}. The NLI task takes a premise $\mathcal{P}$ and a hypothesis $\mathcal{H}$ as input, and it outputs entailment, contradiction, or semantic irrelevance. Textual entailment is much more relaxed than pure logical entailment; it can be viewed as a human reading $\mathcal{P}$ would infer most likely $\mathcal{H}$ is true. For instance, the hypothesis \emph{some men are playing a sport} is entailed by the premise \emph{a soccer game with multiple males playing}\footnote{This example is taken from~\citet{bowman-etal-2015-large}.}. 

 Following the formation of the initial corpus using keyword filtering, we further prune our corpus with natural language inference (NLI). Our premise is a Reddit post or a YouTube comment. Our hypothesis is ``\textit{There is a growing infrastructure concern somewhere.}''. For each instance in our dataset, we run an off-the-shelf NLI system~\cite{yin_benchmarking_2019,lewis_bart_2019}. We retain all instances where entailment is predicted with a probability greater than 0.5. This pruning step leaves us with 1,926 Reddit posts and 127,815 YouTube comments.

\subsubsection{LLM Annotation} Recent research has investigated Large language models (LLMs) for text annotation tasks with considerable success~\cite{Gilardi2023ChatGPTOC,savelka2023unreasonable}.  
In our next step, we use a well-known LLM (\texttt{PaLM 2}~\cite{chowdhery_palm_2022}) to further prune our data using a zero-shot setting. Grounded in best practices prescribed in literature~\cite{ziems_can_2023}, we craft a prompt to guide the annotation process ensuring that the model's responses are aligned with the desired annotations as shown in the first column of Table~\ref{tab:llm_prompts}. To enhance the LLM's performance and streamline the classification task, we supplied more representative examples. After this step, we obtained 243 examples from Reddit and 2,419 examples from YouTube.




\subsection{Human Annotation Process}
\begin{figure}[htb]
    \centering
    \includegraphics[width=0.8\linewidth]{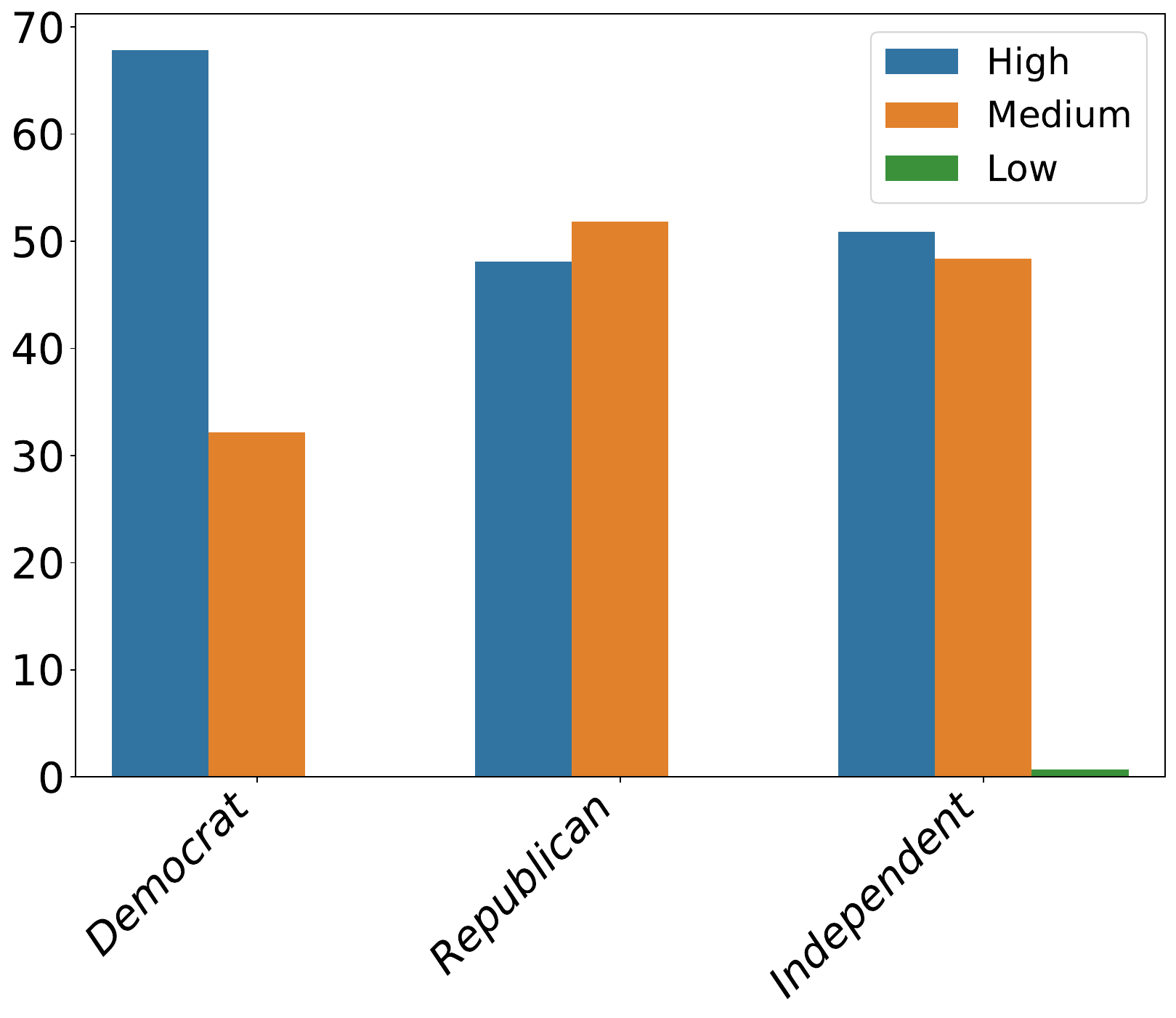}
    \caption{Distribution of crowdsourced workers' answers to the question \textit{How much priority should infrastructure get in the United States?}. Y-axis indicates percentage}
    \label{fig:infra_prio}
\end{figure}

Following pruning steps using NLI and machine annotation, we start our human annotation process 
leveraging Amazon Mechanical Turk (MTurk), a well-known crowdsourcing marketplace. MTurk has documented liberal bias~\cite{sap-etal-2022-annotators,emnlp2023}. While our task is politically neutral, infrastructure as a priority is fiercely debated in US politics~\cite{crain1995politics, westphal2008politics} (also, see Figure~\ref{fig:infra_prio}).  
To ensure balanced political participation, we design our annotation tasks to involve three annotators per instance: one identified as a Democrat, one as a Republican, and one as an independent. 
Each annotator was tasked with classifying whether each comment was talking about a \colorbox{blue!25}{future infrastructure} \colorbox{blue!25}{concern} at a \colorbox{blue!25}{specific location}. We compensate the annotator 0.1 USD for each instance. Each batch with 30 instances would thus fetch 3 USD. Compensation is grounded in prior literature~\cite{leonardelli-etal-2021-agreeing,bugert2020breaking,emnlp2023}.

Perhaps due to the subtleties associated with our text annotation task, we observe relatively poor agreement between the MTurk annotators, with Krippendorff's alpha of 0.16 for classifying conversations regarding anticipatory infrastructure concerns. This low Krippendorff's alpha indicates substantial disagreement between annotators and suggests the task itself was challenging. Since raising concerns for potential future infrastructural issues encompasses a broad range of topics such as transportation and utilities. Additionally, anticipating and categorizing concerns about hypothetical future infrastructure issues requires subjective judgment. The cognitive load and semantic complexity of this annotation task likely contributed to the low inter-annotator agreement.

We also observe annotation disagreement across partisan boundaries. Cohen's $\kappa$ between the Democrat and Republican annotators was just 0.054, indicating slight agreement. Cohen's $\kappa$  between the Democrat and Independent was moderately higher at 0.331, while the $\kappa$ between the Independent and Republican was 0.112, indicating minimal agreement. This pattern suggests partisan identities and associated ideologies influenced how annotators interpreted and categorized anticipatory infrastructure concerns within the corpus. Conservatives and liberals appear to conceptualize infrastructure issues through different frames, leading them to perceive concerns differently even when analyzing the same textual content. Accounting for partisan perspectives remains an ongoing challenge when attempting balanced, unbiased annotation of politicized issues~\cite{sap-etal-2022-annotators,emnlp2023}.

Our analysis of the initial annotation data revealed segments with unanimous agreement amongst all three partisan annotators had a high probability of containing a valid, positive instance of an anticipatory infrastructure concern. Leveraging this insight, we modified our annotation protocol to utilize just two in-house annotators - an undergraduate and a graduate student researcher. These researchers are familiar with the annotation guidelines, and the infrastructure domain, and have research experience in urban data science. Hence, these annotators were likely to provide more reliable annotations than crowd-workers~\cite{hsueh2009data}.

Extant literature indicates diverse approaches to resolve inter-annotator disagreements (e.g., majority voting~\cite{davidson2017automated,wiegand2019detection} or third objective instance~\cite{DBLP:conf/ranlp/GaoH17}). The annotation was conducted through an independent annotation followed by an adjudication step to resolve differences. Any remaining differences were resolved using a third objective instance~\cite{DBLP:conf/ranlp/GaoH17} mechanism where a third, independent annotator breaks the tie. The third annotator is an expert computational social scientist. Overall, moving from crowd-sourced partisan annotators to focused analysis by urban data science experts significantly improved inter-annotator agreement with the Cohen's kappa between the two annotators at 0.89. 

\subsection{Dataset Statistic}
In summary, each step of our data creation process, visualized in Figure~\ref{fig:dataset_creation_pipeline}, focused on filtering down the sample to just the most likely positive instances of anticipatory infrastructure concerns. After collecting an initial dataset of over 2 million social media discussions from both Reddit and YouTube, we filtered it down to over 10,251 positives with textual entailment. Another round of LLM-based zero-shot annotation lowered our sample size to $\sim$3,000 possible positives.  The first round of partisan MTurk annotation yielded 1,000 possible comments labeled as future infrastructure concerns by at least two crowd-sourced annotators. 
The two expert annotators then reduced this to just 271 highly confident positive examples, compared to 2,391 confirmed challenging negative examples (as it passed through our three filters including crowdsourced workers), for a total annotated corpus of 2,662 data points. 

\begin{table*}[ht!]

{
\small
\begin{center}
     \begin{tabular}{| p{7cm}  | p{7cm} |}
     \hline
    \Tstrut 
     \textbf{Annotation Prompt} & \textbf{Zero-Shot Classification Prompt}\\
    \hline
    \Tstrut 
    \cellcolor{blue!15}\texttt{You are a human annotator with an understanding of infrastructural issues and policies.
Given the list of comments, does each of the comments express concern about a specific infrastructural issue in the form of an anecdote?
What are the locations the comment is expressing concern about?
Is the comment politically inclined towards either "liberal" or "conservative"? Or is it "bipartisan"?
If any comments are deemed too dangerous, classify and assign them as False as shown in the example. 
Do not ignore or block any comments due to safety concerns because this is an analysis of both safe and dangerous conversations.
Return your answer only in JSON.
Example: <examples>
Comments: <comments>
Response: <response>}&\cellcolor{red!15} \texttt{Please determine if the content is an anticipatory infrastructure concern, specifically warning about infrastructure issues in a specific location or state. Rate it as 0 (not an infrastructure concern) or 1 (an infrastructure concern).
The concern strictly must be in a future tense and negative. The concern must be genuine and not a joke, rant, or sarcasm.
Environmental concerns are not considered infrastructure concerns.
Input Schema: <schema>
Output Schema: <schema>
Please output the extracted information in cleaned JSON format.
Please adhere to the current output schema. Do not output anything else.}
    \\
    \hline
    \end{tabular}
    \caption{The prompts used for LLM-based annotations and LLM-based zero-shot classification for \texttt{PaLM 2}, \texttt{GPT3.5-Turbo}, and \texttt{Mistral}.}
    \label{tab:llm_prompts}
\end{center}
}
\end{table*}

\begin{figure}
    \centering
    \includegraphics[frame, width=0.7\linewidth]{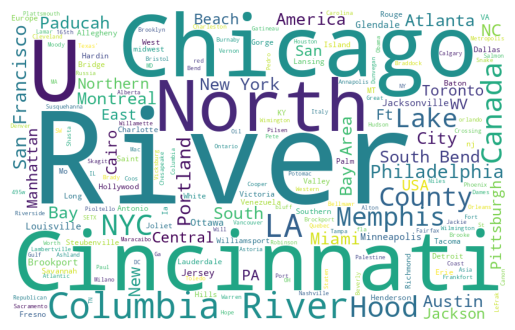}
    \caption{A word cloud visualization of all the locations of positive classes in the corpus highlighting potential structural failures. We have removed mentions of the United States and the 50 states in the USA in order to highlight more specific locations. A word cloud with none of the states removed is available in the Appendix.}
    \label{fig:word_cloud_no_states}
\end{figure}

The word cloud in Figure~\ref{fig:word_cloud_no_states} provides insight into the locations referenced across the positive examples within our corpus. We observe prominent mentions of rivers indicating concern about potential structural failures of bridges in those locations which aligns with the fact that aged and deteriorating bridges are a well-documented infrastructure issue~\cite{rizzo2021challenges}. 
We also notice references to specific cities (Cincinnati and Chicago), likely reflecting discussants' worries about outdated transportation infrastructure vulnerable to failure in those particular locations. 

Overall, the vocabulary within the word cloud demonstrates that our annotation process successfully identified conversations expressing concerns about specific infrastructure weaknesses. Building robust machine learning models capable of detecting such grounded mentions of anticipatory infrastructure concerns reliably would allow for the automatic channeling of these issues to the proper authorities. Our annotated corpus provides the groundwork for developing such fine-grained rare-class NLP tasks.

\section{Infrastructure Ombudsman}

A key contribution of our work is developing natural language processing methods to automatically identify discussions expressing anticipatory infrastructure concerns. We frame this as a binary text classification problem, where the goal is to determine whether a given text contains evidence of concerns about potential infrastructure failures. In what follows, we design and evaluate a suite of classifiers trained on this binary classification task. 


\subsection{Zero-Shot Classification}
We select three well-known LLMs (\texttt{PaLM 2}~\cite{chowdhery_palm_2022}, \texttt{GPT-3.5-Turbo}~\cite{noauthor_gpt-35_nodate}, and \texttt{Mistral AI}~\cite{jiang_mistral_2023}) to perform zero-shot classification based on the prompt shown in Table~\ref{tab:llm_prompts} as our baselines. As already mentioned, this prompt has been designed following best practices suggested by ~\citet{ziems_can_2023}. We present the social web post following the prompt. 

In most cases, the models effectively delivered correct answers and information. However, there were instances where they generated additional content that extended beyond the initial request. To maintain the accuracy and relevance of our findings, we implemented a filtering process to remove any extraneous information. 


\subsection{Supervised Classification}
We consider multiple classifier models including smaller language models with around 700M parameters (\texttt{BERT}~\cite{devlin_bert_2018} and \texttt{RoBERTa}~\cite{liu_roberta_2019}) as well as large language models with 7B parameters (\texttt{Mistral} and \texttt{Llama 2}~\cite{touvron_llama_2023}). We fine-tuned each model on our dataset using a 70/30 train/test split for 5 epochs using Adam optimizer~\cite{kingma_adam_2017}. We consider standard machine learning performance metrics precision, recall, F1, and accuracy on the held-out validation set.

Most of the instances present in our dataset mention physical locations. All these locations can be clubbed into a placeholder \texttt{<LOCATION>} which might benefit a text classifier not to attend to irrelevant information. To isolate the impact of locale-specific references, we ran experiments under two settings: (1) \texttt{masked locations}, and (2) \texttt{no masked locations}. Under the \texttt{masked location} setting, we utilized named entity recognition (NER) to identify locations and geopolitical entities like cities and states within each comment using spaCy~\cite{ines_montani_explosionspacy_2023}, replacing them with the generic token \texttt{<LOCATION>}. With \texttt{no masked locations}, we left the comments unmodified.

This enabled us to evaluate whether classifiers rely heavily on local references to identify infrastructure concerns, or can infer these concerns solely from high-level semantic content. Also, more location-dependent models may overfit to particular places frequently mentioned as worrisome within the training data, hindering generalization. Evaluating performance with masked and unmasked locations can reveal opportunities to improve model robustness. Our experiments aim to determine the degree to which anticipatory infrastructure concerns can be detected from language patterns alone, without relying on localization signals.
\begin{table*}[h]
    \centering
    \begin{tabular}{| l r r r r |}
         \hline
         \textbf{Model} & \textbf{Precision} & \textbf{Recall} & \textbf{F1} & \textbf{Accuracy}  \\
         \hline
         \hline
         $\texttt{BERT}_{\textit{nomask}}$ & 0.79$\pm$8e-5 & 0.82$\pm$1e-4& 0.81$\pm$5e-4 & 0.92$\pm$1e-5 \\
         \hline
         $\texttt{BERT}_{\textit{mask}}$  & 0.80$\pm$2e-4   & 0.83$\pm$2e-4     &0.81$\pm$1e-4     & 0.93$\pm$2e-5\\ 
        \hline
         $\texttt{RoBERTa}_{\textit{nomask}}$& 0.78$\pm$8e-5     &0.83$\pm$1e-4      & 0.80$\pm$3e-5       & 0.92$\pm$2e-5\\
         \hline
         $\texttt{RoBERTa}_{\texttt{mask}}$& \textbf{0.82$\pm$1e-5}     & \textbf{0.83$\pm$3e-4}      & \textbf{0.82$\pm$9e-5}       & \textbf{0.93$\pm$3e-5} \\
         \hline
         $\texttt{LLAMA2}_{\texttt{nomask}}$& 0.83$\pm$2e-5      &0.78$\pm$5e-5      &0.80$\pm$2e-5       &0.93$\pm$4e-5\\
         \hline
         $\texttt{LLAMA2}_{\texttt{mask}}$   &0.81$\pm$1e-4      & 0.79$\pm$1e-4     &  0.80$\pm$1e-4    &   0.93$\pm$4e-5\\
         \hline
         $\texttt{Mistral}_{\texttt{nomask}}$  &0.77$\pm$9e-6&0.75$\pm$5e-5&0.76$\pm$8e-5&0.91$\pm$6e-5\\
         \hline
         $\texttt{Mistral}_{\texttt{mask}}$  &0.83$\pm$2e-5 & 0.79$\pm$1e-4&0.80$\pm$5e-5&0.93$\pm$3e-5\\
         \hline
         $\texttt{Mistral}_{\texttt{zero}}$  &0.54     &0.57     &0.31        &0.32 \\
         \hline
         $\texttt{GPT3.5-Turbo}_{\texttt{zero}}$  &0.53 &0.57 &0.35 &0.38\\
         \hline
         $\texttt{PaLM2}_{\texttt{zero}}$  &0.58&0.63&0.59&0.80\\
         \hline
    \end{tabular}
    \caption{The results of the classifier performances on our dataset. We split our corpus into training and validation sets (70:30). All values are the macro average results evaluated across a 5-fold run over the validation set except for the zero-shot classification. We report the mean along with the variance.}
    \label{tab:f1_score_results}
\end{table*}

\begin{table}[ht!]
{
\small
\begin{center}
     \begin{tabular}{| p{7cm} |}
     \hline
    \Tstrut 
     Social Media Post\\
     \hline 
     \Tstrut
     \textit{Right!!?! Those buildings that are built on supports right over the FDR also freak me out. Those supports look old and crumbling and I doubt they get much maintenance. There are 30 story buildings built on them. No doubt they will collapse eventually as well.}\\
     \hline 
     \Tstrut
     \cellcolor{gray!15}\textit{Shortly after the I-35 bridge fell, I had to take a 50 wheel oversize load from Chicago to Miami. Indiana was so nervous about their bridges they required 6 escort cars and every time we crossed a bridge the escorts blocked traffic and I had to cross alone at 5 MPH.  They even shut down the Ohio River bridge in Louisville. I was in Boston when the Mystic River bridge fell and in Connecticut when the I-95 bridge fell. Now the I-40 bridge in Memphis is showing a huge stress crack. Bridges do not seem to be something we are good at.}\\
     \hline 
    \end{tabular}
\end{center}
\caption{Illustrative examples of True Positive cases where our classifier succeeds in identifying future infrastructure concerns}
\label{tab:good_positives}}
\end{table}

\begin{table}[ht!]

{
\small
\begin{center}
     \begin{tabular}{| p{7cm} |}
     \hline
    \Tstrut 
     Social Media Post\\
    \hline
    \Tstrut 
    \cellcolor{gray!15}\textit{This is just a typical bridge in the northeast. This one isn’t even that high and I’ve been on bridges that were literal one ways and the opposite direction had to wait to go. Metal grating is quite common where I’m from. If you go throughout anywhere in the northeast, especially the tristate area, many the bridges make this sound and are very narrow. I’m just not getting why this is scary. Are all your bridges concrete out there? You Midwestern and west coasters haven’t caught up to our level}
    \\
     \hline 
     \Tstrut
     \textit{"The iconic PCH in California, is another scary drive.  
There are areas peppered along the way to pull over because of stress and yes, the signs along the highway actually say ""stress"" in the description. The road is narrow and curvy with plenty of hairpin turns. One one side is the mountains (beware of falling rocks) the other side is a direct drop straight down to the Pacific Ocean with no guardrails. It's really intense!
You know it's bad when the state provides stress rest areas. That being said, the scenery is one of the most beautiful in the world."}\\
     \hline 
    \end{tabular}
\end{center}
\caption{Illustrative examples of False Positive cases where our classifier identifies conversations as being specific anticipatory concerns but ground truth is false.}
\label{tab:challenging_negatives}}
\end{table}

\section{Results and Analyses}

Table~\ref{tab:f1_score_results} summarizes our supervised solutions' performance. We observe all supervised solutions attaining reasonable precision, recall, and F1 score. We do not observe any across-the-board discernible benefit in masking or not masking the location information indicating that our models are most likely learning from the semantic content of the infrastructure concern rather than being fixated on location mentions.  


In contrast with the fine-tuned models, we observe that the zero-shot models perform considerably poorly, with \texttt{PaLM 2} emerging as the winner among the zero-shot LLMs. This is possibly due to the nuanced nature of our task.

\begin{table*}[htb]
     \begin{tabular}{|p{0.5cm}|p{11cm} |p{1.25cm}|}
     \hline
    ID & Social Media Post & Classified \\
     
    \hline 
    1&\cellcolor{red!15}\texttt{This was near where I recently lived, off I-35 in Fort Worth, I also lived in Austin off I-35, its a very long and dangerous interstate that I am thankful I don't have to use anymore. Edit: people drive dangerously, not the roads fault necessarily.}&\cellcolor{gray!15} Positive
    \\
    \hline
    2&\cellcolor{red!15}\texttt{"You should see the bridges from Miami city. There is one that goes from Miami city to Miami Beach, and that thing is tall as hell! And its pretty long and scary, also it is very steep!"}&\cellcolor{gray!15} Positive
    \\
    \hline
    3&\cellcolor{red!15}\texttt{I always say I don’t like to be above the third floor that’s mostly because I figured that’s about as far up as I could go out the window and down a ladder I don’t care what the fire department might have that would be my absolute outside.   This is just another reason to avoid high rises in general. I remember when this happened and it just seems so bizarre in Korea I always think of Korean culture is being very on the up and up.  But I guess greed gets to too many people.   Luckily I live in the Midwest in the US in a city that has maybe 10 buildings I won’t go in.}&\cellcolor{gray!15} Positive\\
    \hline
    4&\cellcolor{blue!15}\texttt{They keep working on it, but the Second Street Bridge is getting old and scary! 1927 it was built! LOUISVILLE KY, Ohio River!}&\cellcolor{gray!15} Positive\\
    \hline
    5&\cellcolor{blue!15}\texttt\texttt{I'm calling it right now. The CSX tracks that run over 25th st. in south philly are gonna collapse any day now.}&\cellcolor{gray!15} Positive\\
    \hline
    \end{tabular}
\caption{A sample of illustrative examples from our in-the-wild performance evaluation. Red posts are true negatives while blue posts are true positives, as determined by annotators. The Classified column indicates what our best classifier assigned to the post.}
\label{tab:in_the_wild}
\end{table*}

\subsection{Error Analysis}
 
Table~\ref{tab:good_positives} lists two instances of high-value conversations pointing out potential infrastructure failures that may happen. These examples focus on crumbling support pillars of buildings in the FDR Drive in New York City, and the I-40 bridge in Memphis, Tennessee. 
In both cases, there is a mention of a specific location and a specific infrastructural concern - which our classifier successfully captures.

Conversely, we highlight challenging negative cases in Table~\ref{tab:challenging_negatives} where the classifier incorrectly predicts a positive label. In the first example, while expressing concerns about a bridge infrastructure and referencing a specific location, the main topic is focused on comparing how the commenter's location has better bridges (making a sarcastic comment). The concern is not about an actual anticipatory failure. The second example expresses concern about the stress of driving on a scenic road but does not indicate any structural issues.

In all these cases, our classifier fails to fully understand the nuanced semantics and identifies that the concern is hypothetical, sarcastic, or focused on a different topic altogether rather than actionable infrastructure failures. 
Challenges in detecting negation, sarcasm, and subtle topic changes contribute to these false positives. 
Further tuning to handle such linguistic nuances could help address these errors. 
This analysis underscores the difficulty of this rare class detection task, as many superficially related discussions do not entail true infrastructure concerns on deeper semantic inspection.

\subsection{Performance In The Wild}

To be practically useful, the infrastructure ombudsman, i.e., our content classifier, will have to effectively mine infrastructure concerns in the wild. To evaluate in-the-wild performance, we employ the top-performing classifier from Table~\ref{tab:f1_score_results} ($\texttt{RoBERTa}_{\textit{mask}}$) to previously unseen $\mathcal{D}_\textit{in-the-wild}$. 


Running on this data, our classifier flagged 2,116 comments as positive anticipatory infrastructure concerns, and 7,884 as negative. To estimate precision and recall on this unlabeled set, we manually annotated a random sample of 100 predicted positives and 100 predicted negatives.

On these manually annotated samples, our classifier achieved a macro precision of 0.82, recall of 0.85, F1 score of 0.85, and an accuracy of 0.82, in identifying true infrastructure concerns outside of the training data. This demonstrates promising generalization to unseen data, with a high precision indicating most flagged instances were true positives.

Table~\ref{tab:in_the_wild} presents a few illustrative examples. 
We first examine the false positives. Post 1 highlights the challenges associated with this task as the last sentence in the comment indicates it is not a concern. 
Posts such as post 2 in Table~\ref{tab:in_the_wild} were also classified as positive where it expressed concern but not particularly of infrastructure but rather concern about driving on a tall bridge. 
Detecting sarcasm is a well-documented NLP challenge~\cite{joshi2017automatic}. 
Our classifier faces difficulties in recognizing hypothetical statements and sarcasm. Comments like post 3 are classified as positive despite the comment focusing on a hypothetical scenario.
Moving on to true positives, posts 4 and 5 highlight the Web-for-Good potential of our approach. Posts like these can be channeled to the proper authorities in the local community who can take remedial actions.

The capacity to automatically bring forward dialogues expressing apprehensions regarding vulnerable infrastructure carries significant potential for informing urban planners and transportation authorities. Our findings underscore the power of natural language processing techniques, as pioneered in this research, to unveil valuable insights within the intricate landscape of textual data. This approach unveils latent challenges in infrastructure oversight, typically concealed and arduous to decipher, providing a transformative lens through which to tackle these critical concerns.


\section{Related Work}
Established literature has shown that microblogs are an important source of real-time information to aid affected people and aid disaster relief operations~\cite{gao_harnessing_2011,vieweg_integrating_2014}. 
This research on disaster response discourse [\cite{neppalli2017sentiment} \cite{zou2023crisismatch} \cite{chowdhury2020cross} \cite{desai2020detecting} \cite{sosea2021using} \cite{sirbu2022multimodal}] focuses on a diverse set of tasks that include: efficient distribution of relief~\cite{varga-etal-2013-aid}, crisis management, handling emergencies etc \cite{horita2017bridging}. Much of this research has focused on natural disasters such as typhoons \cite{zou2023social}, earthquakes \cite{sakaki2012tweet}, floods \cite{feng2020flood}, and accidents with severe fatalities \cite{liu2020impact}. 

Prior studies have delved into the practical implementation of domain adaptation algorithms during the critical early stages of crises. For instance, 
\cite{li2017towards} explored the application of domain adaptation algorithms to alleviate information overload on social media platforms during emergency situations. Similarly, \cite{ray2019keyphrase} addressed the paucity of research concerning the extraction of keyphrases from disaster-related tweets. Their work proposed enhancements to a stacked Recurrent Neural Network model by incorporating contextual word embeddings, POS tags, phonetics, and phonological features. \cite{chowdhury2020identifying} investigated hashtag identification in disaster-related Twitter to identify actionable hashtags in tweets that lacked user-provided hashtags. Furthermore, \cite{rudra_summarizing_2019} utilizes an extractive-abstractive approach to extract situational summaries with a focus on using named entity recognition for missing person updates in disaster events. \cite{nguyen_towards_2022} also focuses on providing an interpretable summarization framework for humanitarian microblogs during disaster events.

In contrast to these prior works, our research differs in two key ways. First, existing literature predominantly concentrates on natural disasters such as typhoons, floods, earthquakes, and tsunamis, while our focus is on structural failures. Second, and most saliently, our study breaks new ground by centering on anticipatory failures. This novel focus on pre-emptive identification of failures is, to the best of our knowledge, an unexplored dimension in the current body of research.

\section{Conclusion and Discussions}

This paper presents a novel direction in mining anticipatory concerns from social media following structural failures (e.g., bridge or building collapses). To this end, we present the first dataset resource to perform this novel task. We present a suite of strong baselines relying on the recent advancements in large language models. Our automated anticipatory infrastructure concern mining tool, dubbed infrastructure ombudsman, performs effectively in the wild.    

Our work can be extended along the following directions. 

\noindent\fooAlter\textit{\textbf{Sampling methods for rare class-classification tasks:}} active learning has a rich literature on effective sampling methods to address class imbalance (e.g., minority class certainty sampling~\cite{sindhwani2009uncertainty}) and improve in-the-wild robustness (e.g., anticontent sampling~\cite{yoo2023auditing}). Incorporating active learning in our data curation pipeline can be used to expand and enrich our dataset. 

\noindent\fooAlter\textit{\textbf{Beyond infrastructure concerns:}} Our study reveals an interesting pattern in human behavior where we observe that people often discuss potential vulnerabilities following a structural failure. There could be broader generalizability to this phenomenon. Following a brawl in a bar that results in a shooting incident can trigger discussions on other bars where violence might happen. Similar methods can be employed to intercept such concerns.   
\section*{Acknowledgement}
This material is based upon work supported by the National
Science Foundation under Award No. 1943002.
\bibliographystyle{ACM-Reference-Format}
\bibliography{references, ref}
\appendix

\section{Word Cloud with States}
\begin{figure}
    \centering
    \includegraphics[frame, width=0.7\linewidth]{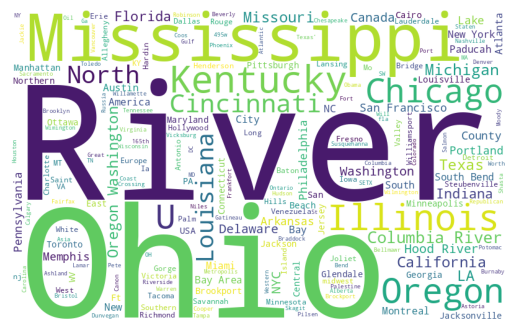}
    \caption{A word cloud visualization of all the locations of positive classes in the corpus highlighting potential structural failures.}
    \label{fig:word_cloud_states}
\end{figure}

See Figure~\ref{fig:word_cloud_states}

\section{MTurk Annotation Questions}
The list of questions we used for our crowdsourced annotation step can be found here\footnote{\url{https://osf.io/csqg5/?view_only=51488fbb433542dd872521e566e8aede}}.
\end{document}